\documentclass[twocolumn,aps,pra,10pt]{revtex4-1}

\usepackage{amsmath}
\usepackage{graphicx}

\newcommand{\sref}[1]{Sec.~\ref{#1}}
\newcommand{\xref}[1]{Eq.~\eqref{#1}}
\newcommand{\Xref}[1]{Equation~\eqref{#1}}
\newcommand{\fref}[1]{Fig.~\ref{#1}}
\newcommand{\Fref}[1]{Figure~\ref{#1}}

\newcommand{\ket}[1]{|{#1}\rangle}
\newcommand{\bra}[1]{\langle{#1}|}
\newcommand{\braket}[2]{\langle{#1}|{#2}\rangle}

\newcommand{\CD}{_\textrm{CD}}
\newcommand{\cor}{_\textrm{corr}}
\newcommand{\sz}{^{(0)}}
\newcommand{\so}{^{(1)}}
\newcommand{\sm}{^{(m)}}
\newcommand{\NN}{{\mathcal N}}
\newcommand{\Hamil}{{\mathcal H}}
\newcommand{\Vamil}{{\mathcal V}}
\newcommand{\Padi}{P_\textrm{adiabatic}}

\begin{document}

\title{Adiabatic theorem revisited: the unexpectedly good performance of adiabatic passage}

\author{Albert Benseny}
\email[e-mail:]{albert.benseny@q-ctrl.com}
\affiliation{Q-CTRL, Sydney, NSW Australia \& Los Angeles, CA USA}

\author{Klaus M{\o}lmer}
\email[e-mail:]{moelmer@phys.au.dk}
\affiliation{Center for Complex Quantum Systems, Department of Physics and Astronomy, Aarhus University, \\ Ny Munkegade 120, DK-8000, Aarhus C, Denmark}

\begin{abstract}
Adiabatic passage employs a slowly varying time-dependent Hamiltonian to control the evolution of a quantum system along the Hamiltonian eigenstates. For processes of finite duration, the exact time evolving state may deviate from the adiabatic eigenstate at intermediate times, but in numerous applications it is observed that this deviation reaches a maximum and then decreases significantly towards the end of the process. We provide a straightforward theoretical explanation for this welcome but often unappreciated fact. Our analysis emphasizes a separate adiabaticity criterion for high fidelity state-to-state transfer and it points to new effective shortcut strategies for near adiabatic dynamics.
\end{abstract}

\maketitle

\begin{figure*}
	\centerline{\includegraphics[width=0.94\textwidth]{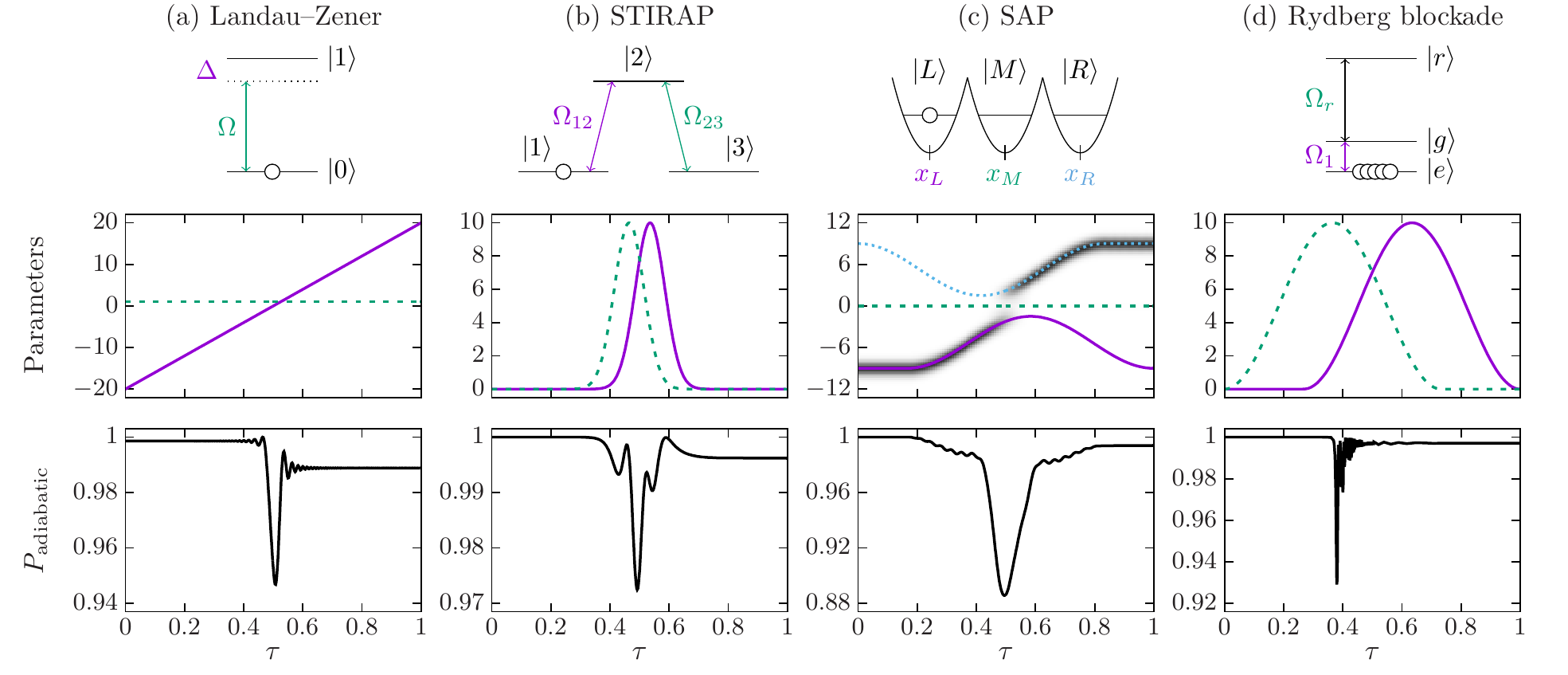}}
\caption{
Examples of application of adiabatic passage in different physical systems.
In column (a), the top figure depicts the energy diagram for the Landau--Zener transition~\cite{Landau,Zener,Stueckelberg,Majorana}, describing, e.g., a frequency chirped laser excitation of a two-level atom.
The lower panels show the time-dependent detuning $\Delta$ (purple solid line) and Rabi frequency $\Omega$ (green dashed line), and the population in the adiabatically followed eigenstate (black line), respectively.
In column (b), the top figure depicts the energy diagram for Stimulated Raman adiabatic passage (STIRAP) in a $\Lambda$ system~\cite{STIRAP1,STIRAP2}.
The lower panels show the Rabi frequencies of the pump $\Omega_{12}$ (purple solid line) and Stokes $\Omega_{23}$ (green dashed line) pulses, and the population in the adiabatically followed eigenstate (black line), respectively.
In column (c), the top figure depicts the triple-well trapping potential to transport a particle via spatial adiabatic passage (SAP)~\cite{Kai2004,SAPreview}.
The lower panels show the three trap center positions (purple solid, green dashed, and blue dotted lines) and the atom density (grey shading), and the population in the adiabatic eigenstate of the Hamiltonian (black line), respectively.
In column (d), the top figure depicts the energy diagram of three-level atoms with an upper Rydberg state causing a Rydberg excitation blockade which permits adiabatic creation of highly-entangled states~\cite{Moller,Abad2018}.
The two lower panels show the Rabi frequencies of the pulses $\Omega_1$ (purple solid line) and $\Omega_r$ (green dashed line), and the population in the adiabatic many-atom eigenstate of 10 atoms.
}
\label{fig:adiabats}
\end{figure*}

\section{Introduction}

A robust and practical method to evolve quantum states employs adiabatic passage where the system at all times occupies an eigenstate of a slowly-varying Hamiltonian that connects the desired initial and final states. Its resilience to variations in the physical parameters makes adiabatic passage ideal for applications with inhomogeneities and slowly-varying perturbations. The usual condition for adiabaticity~\cite{Fock,Messiah} states that the non-adiabatic coupling due to the time dependence of the eigenstate basis should be much weaker than the Bohr frequencies between the energy eigenstates.

\Fref{fig:adiabats} shows near-adiabatic processes in four different systems: the two-level Landau--Zener model~\cite{Landau,Zener,Stueckelberg,Majorana}, the three level Stimulated Raman Adiabatic Passage (STIRAP) process~\cite{STIRAP1,STIRAP2}, wavepacket tunneling dynamics in a time-dependent triple well potential~\cite{Kai2004,SAPreview}, and a multi-atom STIRAP process preparing an entangled state by the presence of Rydberg excitation blockade~\cite{Moller,Abad2018}. In all the cases shown, the system is initialized in an eigenstate of the Hamiltonian, but the system does not exactly follow the time-dependent adiabatic eigenstate during the process. Remarkably, however, towards the end of each process the population, depicted in the lower panels of the figure, returns towards the desired adiabatic eigenstate, $\Padi(t_\textrm{final})\sim 1$. As the time evolution is unitary, it is not possible to remove just any small intermediate deviation from the adiabatic eigenstate by the last part of the process. The lack of adiabaticity in the first part of the process seems to be, almost magically, cancelled by the non-adiabatic evolution in the last part. As we shall show by a simple argument in this article, the system, indeed, follows a very specific trajectory which differs from the adiabatic eigenstate of the time-dependent Hamiltonian, but nevertheless it connects the adiabatic eigenstates in the beginning and the end of the process: ``all's well that ends well''. The purpose of this article is to offer a simple explanation of why this occurs in general in adiabatic processes, and to propose that this insight be exploited more systematically in experiments.

The usual criterion of adiabaticity has been subject to long debate, pointing to its insufficiency, e.g, when weak but temporarily modulated variations of the Hamiltonian are resonant with the energy splitting~\cite{Tong1}. With the precaution of either supplemental criteria on the time derivatives of the Hamiltonian~\cite{Amin,Tong2} or explicit exclusion of resonant oscillations~\cite{Comparat}, the usual criterion does ensure the adiabatic following~\cite{Sarandy,Tong3} to a high degree of precision. Insights from this discussion have also informed the use of adiabatic processes with degenerate Hamiltonians~\cite{Plenio}.

\Fref{fig:adiabats} permits the opposite observation: the usual adiabatic criterion does not necessarily have to be strictly  fulfilled, since the time evolving state may deviate from the adiabatic eigenstate and still find its way back to the adiabatic eigenstate at the end of the process. This is a common observation for many processes and it is in agreement with numerous theoretical analyses. For instance, the Landau--Zener transition model is analytically solvable~\cite{Landau,Zener,Stueckelberg,Majorana} and yields a loss of population that is exponentially small in the process duration, i.e., for a slow process, its dependence on the rate of change of the Hamiltonian and its eigenstates is weaker than any power law. The same exponential suppression has been recovered by more general arguments for a wider range of models~\cite{Dykhne,Davis,Avron,Berry}, and detailed analyses have assessed the important consequences of this favorable behavior for the prospects of adiabatic quantum computing, see, e.g.,~\cite{Jansen,Lidar}.

It is the purpose of the present article to offer a simple and practical explanation of the results shown in \fref{fig:adiabats}. Without embarking on the mathematical analysis leading to the exponential suppression of the non-adiabatic loss of population, we use perturbation theory to demonstrate in a straightforward manner how the time-dependent state can deviate from the adiabatic eigenstates to linear order while the final state deviates to a higher order in the rate of change of the eigenstates. Insights from the same analysis are then used to propose modifications to adiabatic processes, that make them operate at any finite duration.

In \sref{sec:perturbation} we shall apply first order perturbation theory to the equations of evolution in the adiabatic basis and present simple analytical arguments for why the intermediate state error is of first order in the inverse of the process duration while the final state is generally reached with an error that is of a higher order. This result calls the relevance of the conventional adiabaticity criterion into question, and it raises attention to other equally significant criteria for the convergence of the adiabatic passage process.
In \sref{sec:STSA} we shall discuss shortcut-to-adiabaticity strategies which add correcting terms to the Hamiltonian to yield a state evolution that meticulously follows the eigenstate of the originally imposed time-dependent Hamiltonian irrespective of the process duration. We shall show that it is possible to apply these strategies but with weaker correction terms and secure the exact following of higher order, so-called superadiabatic states, which lead to the same perfect final states.
In \sref{sec:conc} we conclude the article, and we comment on the perspectives of the results for different applications of adiabatic passage processes

\section{Perturbation theory of non-adiabatic transitions}
\label{sec:perturbation}

In adiabatic processes, we subject a quantum system to a time-dependent Hamiltonian $\Hamil(t)$, and we aim for the system to follow one of its eigenstates $\Hamil(t) \ket{m\sz(t)} = E_m\sz(t) \ket{m\sz(t)}$ for $t\in [0,T]$. We introduce the rescaled time $\tau = t/T = \epsilon t$ ($T \equiv 1/\epsilon$), exploring the fixed interval $\tau \in [0,1]$, so that the Schr\"odinger equation can be written ($\hbar=1$)
\begin{align} \label{eq:S0}
i \epsilon \partial_\tau \ket{\psi(\tau)} &= \Hamil(\tau) \ket{\psi(\tau)} .
\end{align}
In order to simplify expressions, we will occasionally suppress the $\tau$ dependence in the following.

By expanding the quantum state in the time-dependent adiabatic basis,
$\ket{\psi} = \sum_m c_m\sz \ket{m\sz}$,
the Schr\"odinger equation \eqref{eq:S0} yields equations for the state amplitudes
\begin{align} \label{eq:S0ampl}
i \epsilon \partial_\tau c_n\sz &= E_n\sz c_n\sz
 - i \epsilon \sum_m \braket{n\sz}{\partial_\tau m\sz} c_m\sz,
\end{align}
where the so-called non-adiabatic coupling terms are due to the time evolution of the basis states, and
we use the short hand notation, $\ket{\partial_\tau m\sz} = \partial_\tau\ket{m\sz}$.

The formal structure of the equations of evolution, \xref{eq:S0ampl}, is equivalent to the Schr\"odinger equation with a time-dependent Hamiltonian,
\begin{align}
\label{eq:S1}
i \epsilon \partial_\tau \ket{\psi(\tau)} &= \Big[ \Hamil\sz(\tau) + \epsilon \Vamil\sz(\tau)\Big] \ket{\psi(\tau)},
\end{align}
where
\begin{align}
\label{eq:H_order_0}
\Hamil\sz(\tau) &= \sum_m E_m\sz(\tau) \ket{m\sz}\bra{m\sz},
\\
\label{eq:V_order_0}
\Vamil\sz(\tau) &= \sum_{n,m} V\sz_{n,m}(\tau) \ket{n\sz}\bra{m\sz},
\end{align}
and where we merely regard
\begin{align}
V_{n,m}\sz(\tau) &\equiv - i \braket{n\sz}{\partial_\tau m\sz}
\end{align}
as the matrix elements of a perturbation in a fixed basis $\{\ket{m\sz}\}$.

For a vanishing $\epsilon$, the system follows the adiabatic eigenstates of $\Hamil(\tau)=\Hamil\sz(\tau)$, while the off-diagonal elements of $V\sz_{n,m}(\tau)$ cause transfer of state amplitude of order $\epsilon V\sz_{n,m}/(E_m\sz-E_n\sz)$ among the eigenstates. The diagonal terms $E_m\sz(\tau)$ and $V\sz_{n,n}(\tau)$ impose accumulation of the dynamical phase and the Berry phase~\cite{BerryPhase}, respectively, on the adiabatic eigenstates. We assume a non-degenerate spectrum of $\Hamil$ in which case energy conservation suppresses non-adiabatic transitions. Still, starting from an eigenstate of $\Hamil$ we expect a deviation of the time evolved state from the corresponding time-dependent adiabatic eigenstate of first order in $\epsilon = 1/T$.

\Xref{eq:S1} is the same time-dependent Schr\"odinger equation as \xref{eq:S0}, but by the explicit Eqs.~(\ref{eq:H_order_0},\ref{eq:V_order_0}) for the energy and coupling terms, the time dependent quantum state is represented in the frame of a different, time-dependent basis. It is a key point in the present analysis that, also in this frame, the Hamiltonian terms are slowly evolving and the solution of \xref{eq:S1} may thus adiabatically follow the time-dependent eigenstates $\{ \ket{n\so} \}$ of the Hamiltonian $\Hamil\so(\tau) \equiv \Hamil\sz(\tau) + \epsilon \Vamil\sz(\tau)$.

To explore the non-adiabatic corrections to the evolution along the eigenstates of $\Hamil\so(\tau)$ we exploit the smallness of $\epsilon$, and determine the eigenvalues and eigenstates of the Hamiltonian in \xref{eq:S1} by first order perturbation theory,
\begin{align}
E_n\so(\tau) &= E_n\sz(\tau) + \epsilon \bra{n\sz} \Vamil\sz(\tau) \ket{n\sz} + O(\epsilon^2)
\end{align}
and
\begin{align}
\label{eq:perturbed_states}
\ket{n\so(\tau)} &= \ket{n\sz} + \epsilon \sum_{k\neq n} M\sz_{k,n}(\tau) \ket{k\sz} + O(\epsilon^2)
\end{align}
where we have introduced
\begin{align}
M\sz_{k,n}(\tau) &\equiv \frac{\bra{k\sz} \Vamil\sz(\tau) \ket{n\sz} }{E_n\sz(\tau) - E_k\sz(\tau)}.
\end{align}

Applying these expression in the expansion of the time evolving quantum state, $\ket{\psi(\tau)} = \sum_n c_n\so(\tau) \ket{n\so(\tau)}$, we obtain to second order in $\epsilon$,
\begin{align} \label{eq:amplS2}
i \epsilon \partial_\tau c_n\so(\tau)
&= E_n\so(\tau) c_n\so(\tau) + \epsilon^2 \sum_m V\so_{n,m} c_m\so(\tau)
\end{align}
where we have introduced the coupling terms
\begin{align}
V\so_{n,m} &\equiv
- \frac{i}{\epsilon} \braket{n\so(\tau)}{\partial_\tau m\so(\tau)}
\\ \nonumber
&=
- i
(1-\delta_{n,m}) \partial_\tau M\sz_{n,m} + O(\epsilon).
\end{align}

Analogous to Eqs.~(\ref{eq:S0},\ref{eq:S1}), we can write \xref{eq:amplS2} in the form
\begin{align} \label{eq:pertfin}
i \epsilon \partial_\tau \ket{\psi(\tau)} &= \Big[ \Hamil\so(\tau) + \epsilon^2 \Vamil\so(\tau)\Big] \ket{\psi(\tau)},
\end{align}
where the Hamiltonian operators, describing the dynamics in the frame with the time-dependent basis states $\ket{n\so(\tau)}$, are defined by the matrix element specified above. Note the $\epsilon^2$ factor in \xref{eq:amplS2} and \xref{eq:pertfin} which cause the transfer of amplitude between the adiabatic eigenstates $\ket{n\so(\tau)}$ of $\Hamil\so(\tau)$ to be proportional with $\epsilon^2 = 1/T^2$ rather than with $\epsilon$.

We conclude that the system prepared at $t=0$ in one of the states, $\ket{n\so}$, will follow the time dependence of that eigenstate adiabatically apart from non-adiabatic corrections that are second order in $\epsilon=1/T$. We emphasize that we have not imposed any modification of the time-dependent Hamiltonian $\Hamil$ in the Schr\"odinger equation, \xref{eq:S0}. It is merely due to the use of different time-dependent bases, that \xref{eq:S1} and \xref{eq:pertfin} appear with different instantaneous diagonal parts and different perturbative corrections. Certainly, the basis states $\ket{n\so(\tau)}$ in \xref{eq:perturbed_states} deviate from the adiabatic eigenstates of the original Hamiltonian $\Hamil$ by a linear amount in $\epsilon$, but if we ensure that these deviations vanish at both $t=0$ and $t=T$, the final state will occupy the desired eigenstate $\ket{n\sz}$ of the final Hamiltonian $\Hamil$ with at most a second order correction. While this relaxes the criterion on the smallness of $\epsilon$, it demands that $\ket{n\so(\tau=0)}=\ket{n\sz(\tau=0)}$ and $\ket{n\so(\tau=1)}=\ket{n\sz(\tau=1)}$, which is fulfilled if $\Vamil\sz(\tau)$ vanishes at those times, i.e., if the rate of evolution of the Hamiltonian $\Hamil$ is continuous and goes to zero at the end and beginning of the process.

This analysis explains why first order deviations from the adiabatic eigenstates may readily occur during time evolution, why they disappear, and why the accumulated non-adiabatic correction at the end of the process is easily restricted to be of a higher order, cf. the lower panels in \fref{fig:adiabats}. The reader may also consult the upper panels of column (a) in Figs.~3 and 4, where the solid, purple curve shows the population of the time evolved state on the adiabatic basis state, while the dashed, green curve shows the near-unit population on the corresponding first superadiabatic basis state. The initial and final vanishing of the rate of evolution of the Hamiltonian, and hence of $\Vamil\sz(\tau)$, is important, and the commonly used linear interpolation between a pair of initial and final Hamiltonians, $\Hamil(t) = (1-t/T)\Hamil_A + (t/T)\Hamil_B$, as well as STIRAP laser pulses with truncated Gaussian or sinusoidal shapes, may all introduce first order corrections. Corrections of this order may, however, be readily eliminated by merely tapering these functions to their initial and final values in a more smooth manner. While this is possible only at the expense of leaving a shorter time for the remaining time evolution, the effective increase of $\epsilon$ is easily outweighed by the absence of first order coupling terms in \xref{eq:pertfin}. There is nothing magic about the initial and final times, and if $\Hamil$ smoothly approaches constant values at definite time intervals during a process, the deviation from the instantaneous eigenstate will also be of second order or higher in $\epsilon$ at those times. This result may be related to recent empirical studies of open system dynamics that show that pausing the dynamics may increase the ground state yield in quantum annealers~\cite{LidarPause}.

We note that the realization that the non-adiabatic coupling can be treated as a perturbation on the adiabatically evolving Hamiltonian may be exploited in a more direct, formal transformation without the explicit reference to the first order perturbed eigenstates. Indeed, the Schrieffer--Wolff transformation~\cite{SW} employs a unitary operator $\exp(S)$, where $[\Hamil\sz,S]=\epsilon\Vamil\sz$, to transform $\Hamil\sz + \epsilon\Vamil\sz$ into the Hamiltonian $\tilde{\Hamil} = \Hamil\sz + [S,\epsilon\Vamil\sz]/2 + [S,[S,\epsilon\Vamil\sz]]/3 ... $, which is at least second order in the perturbation  as $S$ is of the same order as $\epsilon\Vamil\sz$.

\section{Shortcut to superadiabaticity}
\label{sec:STSA}

For many quantum state control problems, the action of dissipation, loss and external perturbations may be at variance with the time needed to ensure reliable state transfer by adiabatic protocols. This has inspired efforts to find strategies to correct and counteract the non-adiabatic couplings while maintaining a finite duration of the process.
Indeed, it is possible to cancel these couplings and follow the eigenstates $\{ \ket{n\sz} \}$ of $\Hamil$ exactly by adding the following counterdiabatic term to the Hamiltonian~\cite{Berry09}
\begin{equation} \label{eq:HCD}
\Hamil\CD
 = i \epsilon \sum_{n} \Big(\ket{\partial_\tau n\sz}\bra{n\sz} - \braket{n\sz}{\partial_\tau n\sz} \ket{n\sz}\bra{n\sz} \Big)
\end{equation}

The first term in $\Hamil\CD$ exactly cancels the term $\epsilon \Vamil\sz$ in \xref{eq:S1} and is sufficient to suppress the transitions between the eigenstates. It is customary, however, to include the second term, since this will lead to the Berry phase evolution, which is of geometric character, and herewith the shortcut dynamics acquires the same value as under perfect adiabatic evolution.
Application of $\Hamil+\Hamil\CD$ leads to so-called transitionless quantum driving~\cite{Berry09} and has inspired the development of the rich field of shortcuts to adiabaticity~\cite{Torrentegui,Odelin}. If $\Hamil\CD$ can be implemented exactly, or approximately~\cite{Opatrny1,Opatrny2}, this may provide good laboratory solutions for controlled quantum dynamics, and in theory studies it may be combined with optimal control theory~\cite{Henrik}, to ensure unit fidelity while minimizing cost functions representing, e.g., dissipation and experimental limitations.

The cancellation of the non-adiabatic terms is exact, but one of the challenges of transitionless driving is to provide the required strength of the counterdiabatic Hamiltonian. Our demonstration in the previous section that the true evolution of our quantum state differs from a perturbed basis state $\ket{n\so(\tau)}$ to second order in $\epsilon$ suggests to employ a counterdiabatic Hamiltonian to remove the time evolution with respect to that basis, i.e., to establish the transitionless driving in the $\ket{n\so}$ rather than the $\ket{n\sz}$ basis. For a classical analogy, a rider on a bobsleigh race track may apply sideway forces to always remain at the bottom of the track, the (adiabatic) equilibrium position at vanishing forward speed. At finite speed, the rider thus fights the tendency to climb the inclined outer barrier at every turn of the race track. Allowing that motion to linear order in the forward speed and applying only the sideway forces to correct higher order dependencies may clearly be a much less strenuous task for the rider, who may still have ample time to reach the bottom curve in the final linear section of the race track.

Since we aim to make the quantum evolution follow a given basis state exactly, the perturbative expressions in \xref{eq:perturbed_states} do not suffice, and we follow instead the so-called superadiabatic bases~\cite{Berry}. These are formal series expansions of orthonormal basis states $\ket{\lambda_n\sm(\tau)}$ of maximum power $\epsilon^m$, and they are followed adiabatically by solutions to the time-dependent Schr\"odinger equation up to a correction of order $\epsilon^{m+1}$.

\begin{figure}	
\centerline{\includegraphics[width=0.46\textwidth]{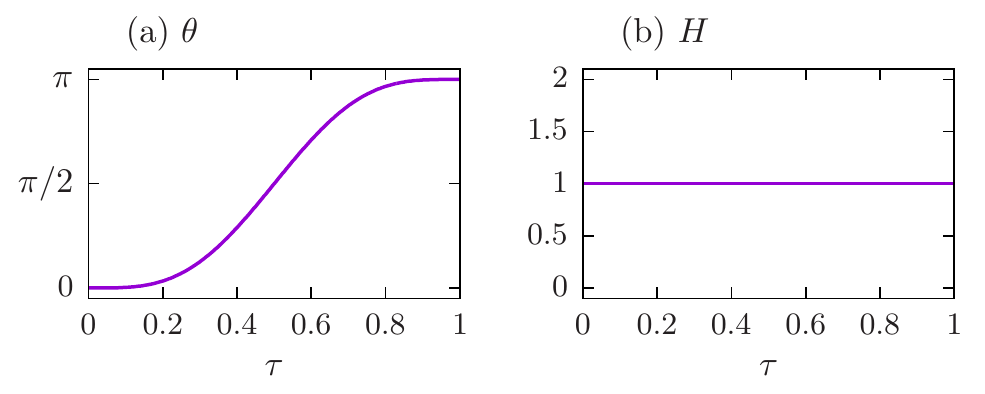}}
\caption{
Parameters for the numerical simulations.
(a) Sweep of the mixing angle $\theta(\tau)$ between 0 and $\pi/2$.
(b) Constant value of $H(\tau) = 1$.
}
\label{fig:paramsT}
\end{figure}

\begin{figure*}
\centerline{\includegraphics[width=0.94\textwidth]{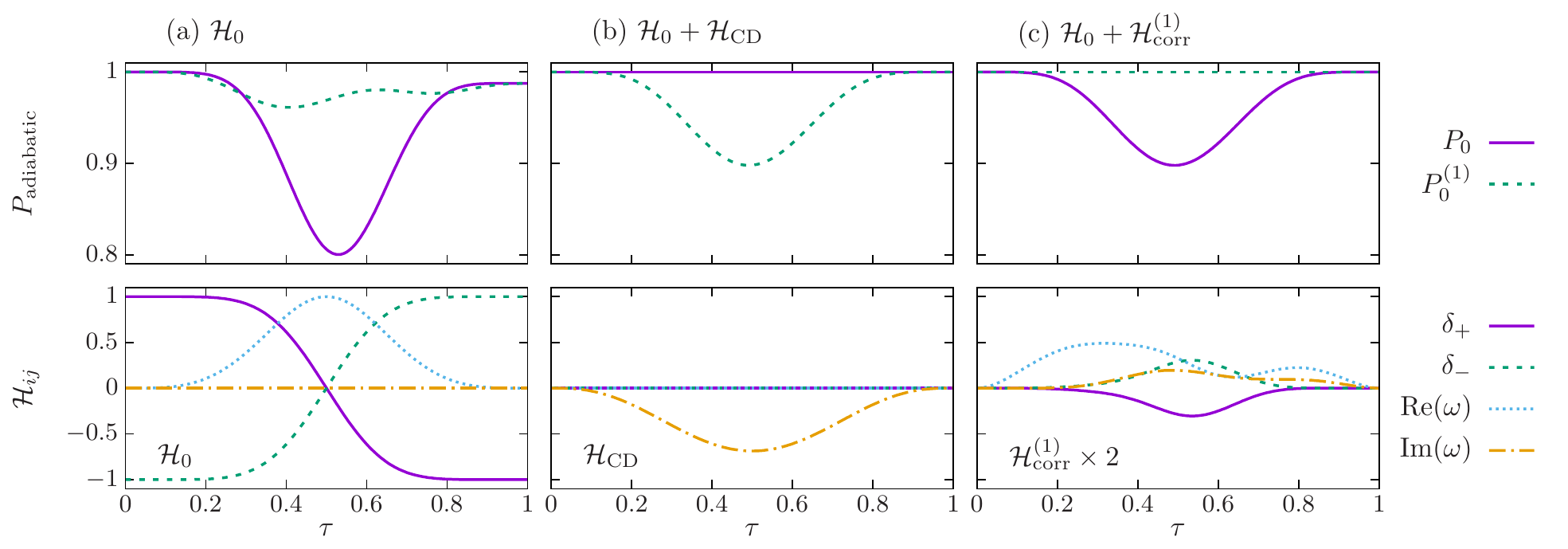}}
\caption{
Comparison of the adiabatic following of different Hamiltonians with the parameters in \fref{fig:paramsT} and for $\epsilon = 0.2$.
(a) Solution for the Hamiltonian $\Hamil_0$.
(b) Solution for $\Hamil_0$ with the counterdiabatic term $\Hamil\CD$.
(c) Solution for $\Hamil_0$ with the counterdiabatic term $\Hamil\cor\so$.
The top row represents the projection on the adiabatic (purple solid line) and first superadiabatic (green dashed line) basis states.
The bottom row shows the elements of the respective Hamiltonians written as
$\Hamil = \left(\begin{smallmatrix} \delta_+ & \omega \\ \omega & \delta_- \end{smallmatrix}\right)$.
For clarity, in (b-c) we show only the perturbation on top of $\Hamil_0$.
}
\label{fig:sims_0.2T}
\end{figure*}

We illustrate this scheme with the example of a spin 1/2 particle subject to a magnetic field with constant magnitude and a direction that sweeps from $z$ to the $x$ direction~\cite{Berry}. The Hamiltonian can be written as the $2\times 2$ system matrix
\begin{align} \label{eq:twolevel}
\Hamil_0 = H \begin{pmatrix} \cos \theta & \sin \theta \\ \sin \theta & -\cos \theta
\end{pmatrix},
\end{align}
where $H$ is a constant, and $\theta(\tau)$ is swept between 0 and $\pi$ for $\tau\in[0,1]$. A symmetric function, $\theta(1-\tau)=\pi-\theta(\tau)$, with $\theta(0) = 0$ and $\theta(1) = \pi$, and vanishing derivatives until the third order for $\tau=0,1$ is given by the function,
$\theta(\tau) = \pi (35 \tau^4 - 84 \tau^5 + 70 \tau^6 - 20 \tau^7)$, shown in \fref{fig:paramsT}(a,b).

$\Hamil_0$ has the eigenvalues $\pm H$ and the eigenstates
\begin{align}
\ket{\lambda_+} = \begin{pmatrix} \cos \theta/2 \\ \sin \theta/2 \end{pmatrix} ,
\quad
\ket{\lambda_-} = \begin{pmatrix} \sin \theta/2 \\ -\cos \theta/2 \end{pmatrix}.
\end{align}
The time derivatives of the eigenstates, $\ket{\partial_\tau\lambda_\pm} = \mp (\dot\theta/2) \ket{\lambda_\mp}$, yield the magnitude $\epsilon\dot\theta/2$ of the counterdiabatic driving Hamiltonian \eqref{eq:HCD}.

Here we shall consider transitionless driving with respect to the first superadiabatic basis, i.e., in contrast to \xref{eq:HCD} which is linear in $\epsilon$, and we shall provide a counterdiabatic driving Hamiltonian of strength $\propto \epsilon^2$ that ensures exact transfer between the initial and final eigenstate of $\Hamil$.
Following Ref.~\cite{Berry}, the first order superadiabatic basis states can be written as
\begin{align}
\ket{\lambda_-\so} &= \NN \Big[ i \epsilon A_1 \ket{\lambda_+} + (1 + i \epsilon B_1) \ket{\lambda_-} \Big]
\\
\ket{\lambda_+\so} &= \NN \Big[ (1 - i \epsilon B_1) \ket{\lambda_+} + i \epsilon A_1 \ket{\lambda_-} \Big]
\end{align}
where
$\NN^{-1} = \sqrt{1+\epsilon^2(A_1^2+B_1^2)}$,
$B_1 = \int (\dot\theta^2/8H) d\tau'$, and
$A_1 = \dot\theta/4H = 2 \dot B_1 / \dot\theta$.
To first order in $\epsilon$, these states are in agreement with the perturbed basis in \xref{eq:perturbed_states}, and we expect that it only requires a Hamiltonian correction of the order of $\epsilon^2$ to follow these states exactly. We evaluate the negative of the non-adiabatic terms, $-i\epsilon\braket{\lambda_n\so}{\partial_\tau \lambda_m\so}$, and we obtain the counterdiabatic Hamiltonian explicitly in the same fixed basis as $\Hamil_0$ \eqref{eq:twolevel},
\begin{align}
\Hamil\cor\so
= \begin{pmatrix}
\delta & \omega \\ \omega^* & -\delta
\end{pmatrix}
\end{align}
where
\begin{align}
\delta &=
\epsilon^2 \NN^2 \Big[
(\cos\theta A_1 - \sin\theta B_1) \dot\theta/2
- \sin\theta \dot A_1
\Big]
\nonumber \\ & \quad
- \epsilon^4 \NN^4 A_1^2 \dot\theta/2 ( 2 A_1 \cos\theta + 2 B_1 \sin\theta )
\\
\omega &=
\epsilon^2 \NN^2 \Big[
\cos\theta \dot A_1 +
\big(B_1 \cos\theta + \NN^2 A_1 \sin\theta \big) \dot\theta/2 \Big]
\nonumber \\ & \quad
+ i \epsilon^3 \NN^2 \Big[ \NN^2 A_1^2 \dot\theta - B_1^2 \dot\theta / 2 - \dot A_1 B_1 \Big]
\nonumber \\ & \quad
+ \epsilon^4 \NN^4
\dot\theta A_1 (\cos\theta A_1 B_1 + \sin\theta (B_1^2 - A_1^2)/2 ).
\end{align}
These results verify that with respect to the first superadiabatic basis, the counterdiabatic Hamiltonian is of second (and higher) order in $\epsilon$.

\begin{figure*}
\centerline{\includegraphics[width=0.94\textwidth]{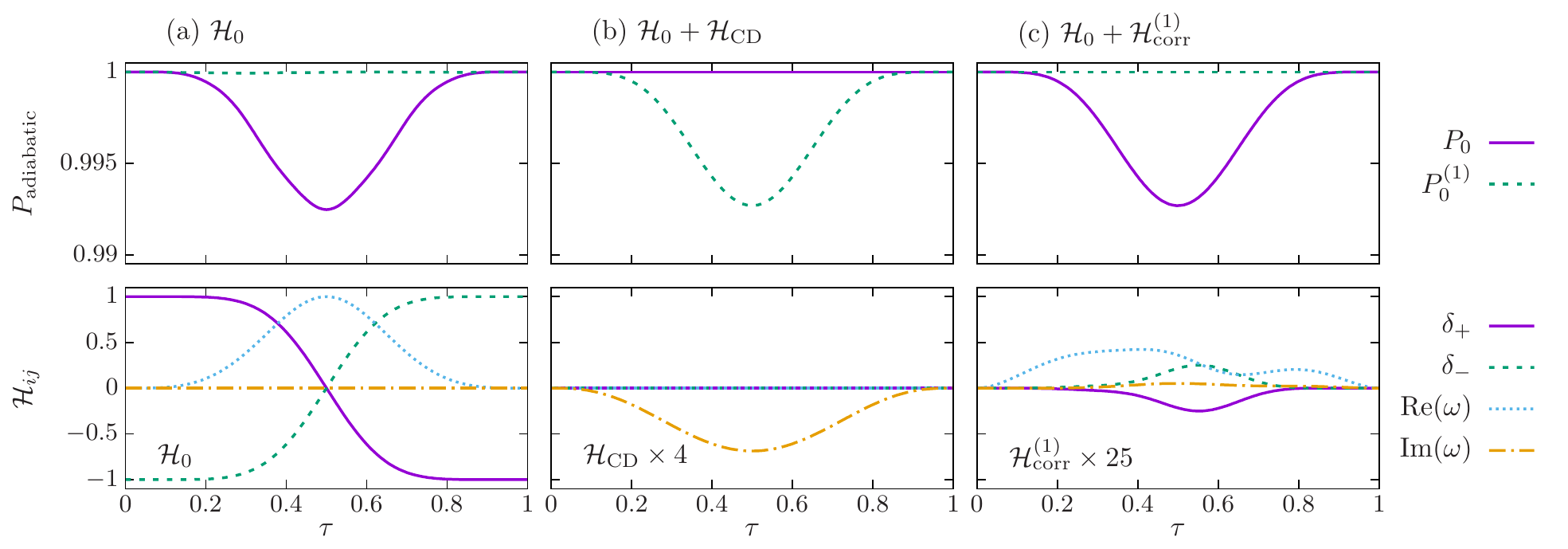}}
\caption{
Comparison of the adiabatic following of different Hamiltonians.
Same as \fref{fig:sims_0.2T}, but for $\epsilon = 0.05$.
}
\label{fig:sims_0.05T}
\end{figure*}

In \fref{fig:sims_0.2T}, we illustrate application of the conventional counterdiabatic Hamiltonian and the counterdiabatic Hamiltonian with respect to the first superadiabatic basis for the evolution of a spin subject to a rotating magnetic field. Time and energy are given in dimensionless units, and the rate parameter is $\epsilon=0.2$, corresponding to a total duration $T=5/H$. The upper plots show the populations of the adiabatic eigenstates (solid purple curve) and of the first superadiabatic basis state (dashed green curve). The lower plots show the diagonal elements of the Hamiltonian as solid purple and dashed green curves, and the real(imaginary) part of the off-diagonal elements of the Hamiltonian as dotted blue (dot-dashed orange) curves. In column (a) we show the populations when the system is subject to only the original Hamiltonian $\Hamil(\tau)$, and we see how the population decreases to 80 \% and recovers to about 98 \% in the adiabatic basis, in analogy with the lower plots in \fref{fig:adiabats}. In the first superadiabatic basis, the intermediate loss of population is smaller, while reaching the same final value. In column (b) we apply the counterdiabatic Hamiltonian terms shown in the lower plot in the basis of the adiabatic eigenstates, to ensure perfect following of an adiabatic eigenstate of $\Hamil(\tau)$, cf. the purple curve in the upper panel. In column (c), we apply the counterdiabatic Hamiltonian terms shown in the lower plot in the original spin eigenstate basis, to ensure perfect following of the first superadiabatic basis state (green dashed line). We note that as expected it requires a weaker correction Hamiltonian to follow the superadiabatic basis state.

We have repeated the calculations with the parameter $\epsilon = 0.05$ and we show the corresponding results in \fref{fig:sims_0.05T}. Due to the smaller value of $\epsilon$ the process is almost adiabatic, and compared to \fref{fig:sims_0.2T}, it takes a four times weaker counterdiabatic Hamiltonian to ensure exact following of an adiabatic eigenstate of $\Hamil(\tau)$, while an approximately 13 times weaker correction is required to ensure perfect following of the first superadiabatic basis state than for $\epsilon = 0.2$.

\section{Conclusion}
\label{sec:conc}

In this article we have provided a simple explanation of why adiabatic passage processes seem to repair deviations from the adiabatic eigenstates accumulated when a system is subject to a Hamiltonian that changes slowly but in a finite time $T$. The time evolving quantum state deviates only to second or higher order from a perturbed state that differs, indeed, to first order from the adiabatic eigenstates of the Hamiltonian, but that state can be proven to connect the initial and final eigenstates exactly if only the Hamiltonian changes smoothly around $t=0$ and $t=T$. In fact, also the second and higher order deviations can be progressively suppressed by comparing with a suitable progression of time-dependent superadiabatic states which may all match the initial and final states exactly. Note that we are not modifying the Hamiltonian for this to happen: the transformed bases in which the dynamics get more and more suppressed are merely introduced for the analysis.

The superadiabatic bases lead to a power series expansion of the error, where coefficients at all finite orders but the last one vanish. This might seem a disturbing fact, but it is a confirmation of the exponentially small error found in Refs.~\cite{Dykhne,Davis,Avron,Berry}. Indeed, the derivatives to any order with respect to $\epsilon$ of $P=\exp(-a/\epsilon)$ vanish as $\epsilon \rightarrow 0$. For any finite value of $\epsilon$, Berry and Lim have shown~\cite{BerryLim}, see also \cite{BerryExp} that due to an asymptotic divergence of the expansion of the superadiabatic basis states, a definite finite order yields the closest agreement with the exact time-dependent solution and closer scrutiny of the argument can be used to derive the exponential suppression of the error. See also a numerical study of the STIRAP process with similar quantitative conclusions~\cite{Elk}.

Even though the formal vanishing of the error to all orders does not warrant a vanishing final state error, our analysis signifies a potential for application of adiabatic processes that goes far beyond the usual adiabaticity criterion. Applications of adiabatic transitions in quantum information science strive for extremely small errors, and the difference between an amplitude error of $\sim \epsilon$ and $\sim \epsilon^2$ can readily become of significance. This implies that already our perturbation theory for the lowest order corrections to the adiabatic dynamics may suffice to guide experiments in a quantitative sense.

We recalled the strategy of shortcuts to adiabaticity which ensures exact following of the adiabatic eigenstates by supplementing the Hamiltonian with counterdiabatic terms.
As a consequence of our analysis we suggest to avoid the effort of maintaining exact evolution of the system along the adiabatic eigenstates and instead apply driving terms that counter only the deviations from the first superadiabatic states. The first superadiabatic state connects the desired initial and final eigenstates of the Hamiltonian. This compensation may be carried out at an even higher order by use of the corresponding superadiabatic bases. The evaluation of the appropriate counterdiabatic Hamiltonian, however, becomes increasingly complicated and the higher order in $\epsilon$ does not guarantee a weaker interaction if its prefactor in the Hamiltonian grows. From a pragmatic point of view, in the limit of small $\epsilon$ and already pretty good adiabatic evolution, the reduction of the strength of the counterdiabatic terms to merely follow the first superadiabatic state may already yield a significant advantage.

Adiabatic preparation of complex many-body states suffers from poor a priori scaling with the system size, but for such systems delicate schemes to permit faster preparation of the full system have been proposed that exploit adiabaticity sequentially on subsets of particles~\cite{YiminGe}, or which apply so-called variational quantum adiabatic algorithms to optimize the ground state overlap~\cite{Schiffer}.
We believe that combination of such ideas with the insights of the present article may hold potential for further progress in the execution of quantum simulations on near-term devices~\cite{NISQ}.

We have presented our analysis for quantum state evolution, but we recall that adiabatic processes have classical counterparts that may show the same phenomena, cf. our analogy with the bobsleigh rider. The fact that the initial-to-final state transfer may be nearly perfect while the intermediate state is not, presents practical possibilities to allow speed-up of a variety of processes at low cost on the final result. This would for example apply to the transfer of classical field amplitudes between wave guides by couplings equivalent to the STIRAP process~\cite{STIRAP1}, and it may apply to the processes in cyclic microscopic and quantum heat engines.

\begin{acknowledgments}
This research was supported by the Villum Foundation and the Danish National Research Foundation through the Grant Agreement DNRF156.
\end{acknowledgments}

\end{document}